\documentclass[12pt,a4paper]{article}
\usepackage[utf8]{inputenc}
\usepackage[english]{babel}
\usepackage{indentfirst}
\usepackage{misccorr}
\usepackage{graphicx}
\usepackage{adjustbox}
\usepackage{amsmath}
\usepackage{amsthm}
\usepackage{caption}   
\usepackage{graphicx}  
\usepackage{caption}
\usepackage{ulem}
\usepackage[labelsep=period]{caption}
\usepackage[labelfont=it]{caption}
\usepackage{makecell}
\usepackage{cite}
\usepackage{xcolor}
\title{\bf Acute ischemic stroke lesion
segmentation in non-contrast CT images using 3D convolutional neural networks}

\author{A.V.~Dobshik$^1$, S.K. Verbitskiy$^1$, I.A.~Pestunov$^2$,  K.M.~Sherman$^3$, \\ Yu.N.~Sinyavskiy$^2$, A.A.~Tulupov$^3$  , V.B.~Berikov$^{1,4}$
\\$^1$Novosibirsk State University, Novosibirsk, Russia
\\$^2$Federal Research Center for Information and \\Computational Technologies, Novosibirsk, Russia
\\$^3$International tomography center SB~RAS, Novosibirsk, Russia\\ $^4$Sobolev Institute of mathematics SB~RAS, Novosibirsk, Russia}
\date{}
\begin{document}
\maketitle

\begin{center}
    \textbf{Abstract}
\end{center}

In this paper, an automatic algorithm aimed at volumetric segmentation of acute ischemic stroke lesion in non-contrast computed tomography brain 3D images is proposed. Our deep-learning approach is based on the popular 3D U-Net convolutional neural network architecture, which was modified by adding the squeeze-and-excitation blocks and residual connections. Robust pre-processing methods were implemented to improve the segmentation accuracy. Moreover, a specific patches sampling strategy was used to address the large size of medical images, to smooth out the effect of the class imbalance problem and to stabilize neural network training. All experiments were performed using five-fold cross-validation on the dataset containing non-contrast computed tomography volumetric brain scans of 81 patients diagnosed with acute ischemic stroke. Two radiology experts manually segmented images independently and then verified the labeling results for inconsistencies. The quantitative results of the proposed algorithm and obtained segmentation were measured by the Dice similarity coefficient, sensitivity, specificity and precision metrics. Our proposed model achieves an average Dice of $0.628\pm0.033$, sensitivity of $0.699\pm0.039$, specificity of $0.9965\pm0.0016$ and precision of $0.619\pm0.036$, showing promising segmentation results.


\underline{\it Citation}: Dobshik~AV, Verbitskiy~SK, Pestunov~IA, Sherman~KM, Sinyav\-s\-kiy~YuN, Tulupov~AA, Beri\-kov~VB, Acute ischemic stroke lesion segmentation in non-contrast CT images using 3D convolutional neural networks. Computer Optics 20XX; 4X(X): XXX-YYY. DOI: 10.18287/2412-6179-CO-editorial index.

\begin{center}
\textbf{Introduction}
\end{center}

Acute cerebral circulation disorder or stroke is a disease with high rates of morbidity and mortality worldwide. According to the American Heart Association, the most common type of stroke is ischemic~\cite{1}. Early diagnosis of stroke is crucial for treatment choice~\cite{2,3}, because tissue changes in the ischemic penumbra may be reversible, especially in the early stages~\cite{4,5}. The choice of diagnostic methods in each specific case strongly depends not only on its applicability (availability, contraindications, patient's condition, etc.), but also on the time of symptoms onset~\cite{6}. Any delay in medical care increases the risk of severe consequences and death.

Neuroimaging is fundamental to most modern methods of differential diagnosis of acute stroke~\cite{5,7,8,9,10,11,12}. For some imaging procedures, contrast injection is required, which is related to the risk of complications and a number of contraindications. The analysis of Computed Tomography (CT) and Magnetic Resonance Imaging (MRI) scans is an integral part of the examination guidelines for patients with signs and symptoms of acute stroke~\cite{7,8,9}.

CT is the most common diagnostic tool for acute  ischemic stroke (AIS) due to its availability (a large number of screening centers, low cost, no contraindications and low body burden) and short screening duration~\cite{7,8,13}. Non-contrast CT (NCCT) was first used in the evaluation of AIS patients in 1995 to detect intracranial hemorrhage (hemorrhagic stroke) and select a treatment strategy at an early time window (within 3~hours of symptoms onset)~\cite{15,16} and proved to be efficient~\cite{13,14,17}. When diagnosing stroke, NCCT demonstrates relatively high specificity and low sensitivity, however, it allows detecting blood clots in cerebral vessels and intracranial hemorrhage (hemorrhagic stroke) which are absolute contraindications for some strategies of AIS treatment~\cite{9,17}.

Interpretation of NCCT scans is associated with certain difficulties, since early AIS changes look like areas of slightly reduced density, which the human eye due to various factors cannot always distinguish from the normal tissues~\cite{3,7,19}. In addition, images often show artefacts (caused by patient movements or imaging camera) that may look like strokes~\cite{8}. Therefore, developing of automated procedures for localization and assessment of AIS tissues volume based on NCCT scans~\cite{19,20,21,22,23,24,25,26} (including the procedures based on convolutional neural networks (CNNs)~\cite{19,20,21,22}) is an urgent task. Randomized controlled trials~\cite{19,27,28,29} have shown that CNN-based methods are comparable to radiologists in terms of sensitivity, specificity, overall accuracy, AUC, ROC, etc. It allows using CNNs as auxiliary tool in clinical practice. In addition, automated processing based on 3D~CNNs allows volumetric analysis taking into account the spatial context and the rapid detection of small ischemic foci.

Over recent years in the medical image segmentation task CNNs have achieved state-of-the-art results~\cite{30,31}. It is due to the convolution operation and its weights capable of learning complicated structures and patterns at multiple scales in the data. The U-shaped encoder-decoder CNN architecture 2D~U-Net~\cite{32} has a wide application~\cite{33,34,35}. In~\cite{33}, a self-adapting framework where 2D~U-Net is adopted to segment various organs is proposed. For the analysis of volumetric images, 3D~CNNs were introduced~\cite{36,37,38}. 3D~U-Net~\cite{36} shows better performance in comparison with 2D~U-Net. In~\cite{39}, a slightly modified 3D~U-Net is utilized for brain tumor segmentation in MRI scans. In~\cite{40} DeepMedic CNN architecture~\cite{38} is used for stroke lesions segmentation and post-processing techniques are investigated. In~\cite{33}, cascaded 3D~U-Net is proposed, which overcomes the disadvantage of 3D~U-Net for datasets with large image sizes, but it requires training two neural networks. U-Net is often improved for a specific task using various architecture choices (e.\,g., Squeeze-and-Excitation blocks~\cite{34,41,42,43}, attention mechanisms, and computer vision transformers~\cite{44}).

In this work, we present a neural network algorithm for the volumetric segmentation of acute ischemic stroke lesions on NCCT brain images. We use a 3D~CNN based on 3D~U-Net architecture~\cite{36}, which we modify with residual connections and relatively new Squeeze-and-Excitation blocks~\cite{41}. To achieve better results, we implement robust pre-processing techniques, a particular patch extraction strategy, and a weighted loss function, which are aimed to alleviate the problem of class imbalance. We perform five-fold cross-validation and compare the results of experiments by measuring Dice, sensitivity, specificity, and precision metrics. 

In the rest of the paper, we describe the used dataset and the method (including data pre-processing techniques and patches sampling strategy), the neural network architecture, and training and testing procedures. In the end, we summarize our work and discuss future plans.

\begin{center}
\textbf{1. Materials}
\end{center}

The dataset used for our study contains volumetric non-contrast CT head scans of 81 patients diagnosed with acute ischemic stroke. The CT images were made by the Philips Ingenuity CT scanner and stored in medical DICOM format. The data were acquired from the International Tomography Center SB~RAS. All volumetric images have the same resolution of $512\times512$, but a different number of slices varying from 306 to 505, depending on the patient. For all images, slice thickness and spacing between slices are 1 and 0.5~mm, respectively. Also, the DICOM attribute pixel spacing, that is, the physical distance in mm between pixel centers, ranges from 0.38 to 0.5 for different volumes. For each volumetric image, corresponding manual segmentation is available. All segmentations were performed by two radiology experts (specialists in magnetic resonance imaging and X-ray computed tomography with 9-13 years experience, PhD in Radiology and Radiation Therapy) using 3D~Slicer~\cite{45}. It is worth mentioning that the number of voxels corresponding to the area affected by acute ischemic stroke is 0.8\% of the total number of voxels in our dataset.

\begin{center}
\textbf{2. Methods}
\end{center}

\begin{center}
\textbf{2.1. Data pre-processing}
\end{center}

\begin{figure}
    \centering
    \begin{tabular}{cc}
        \includegraphics[scale=0.1]{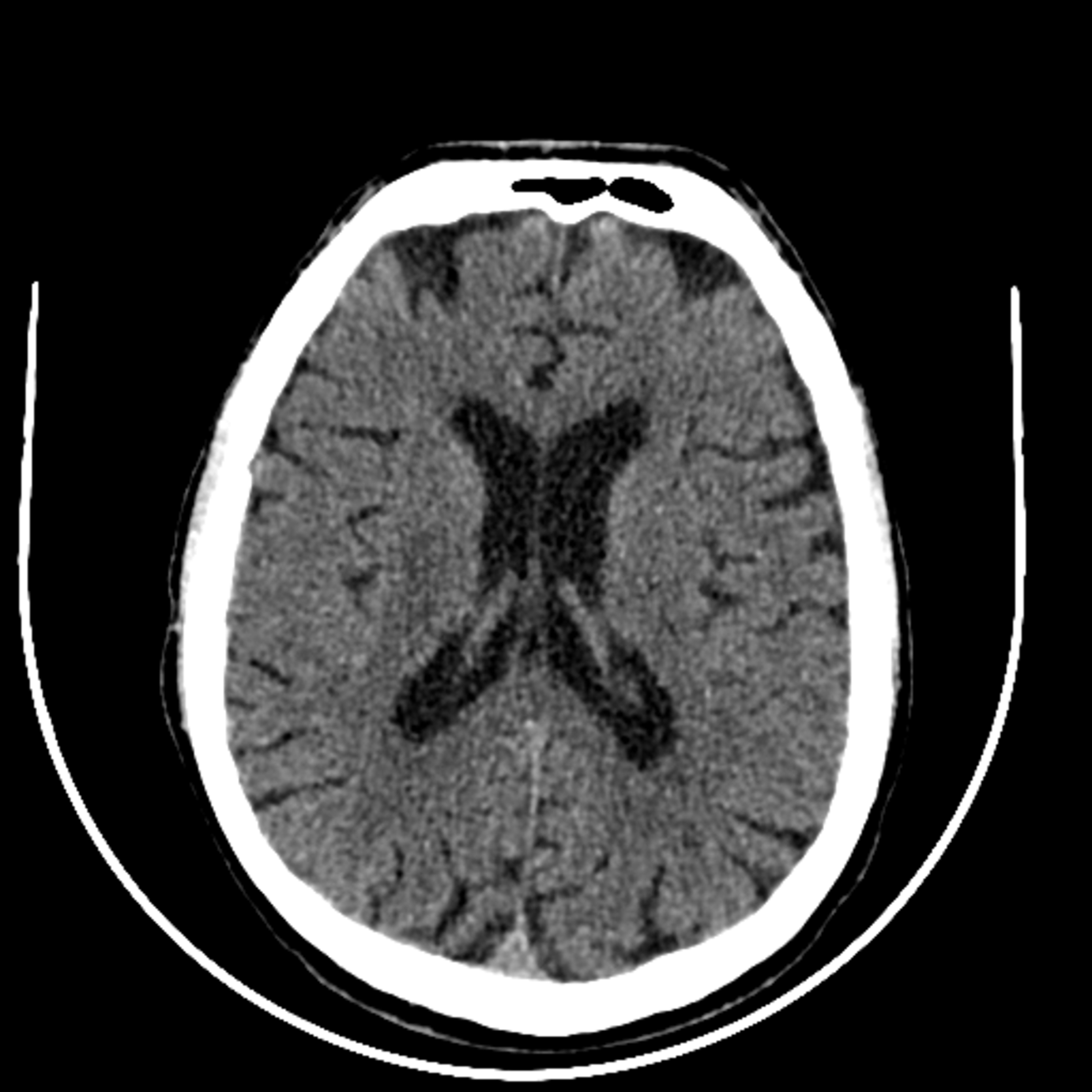} & \includegraphics[scale=0.1]{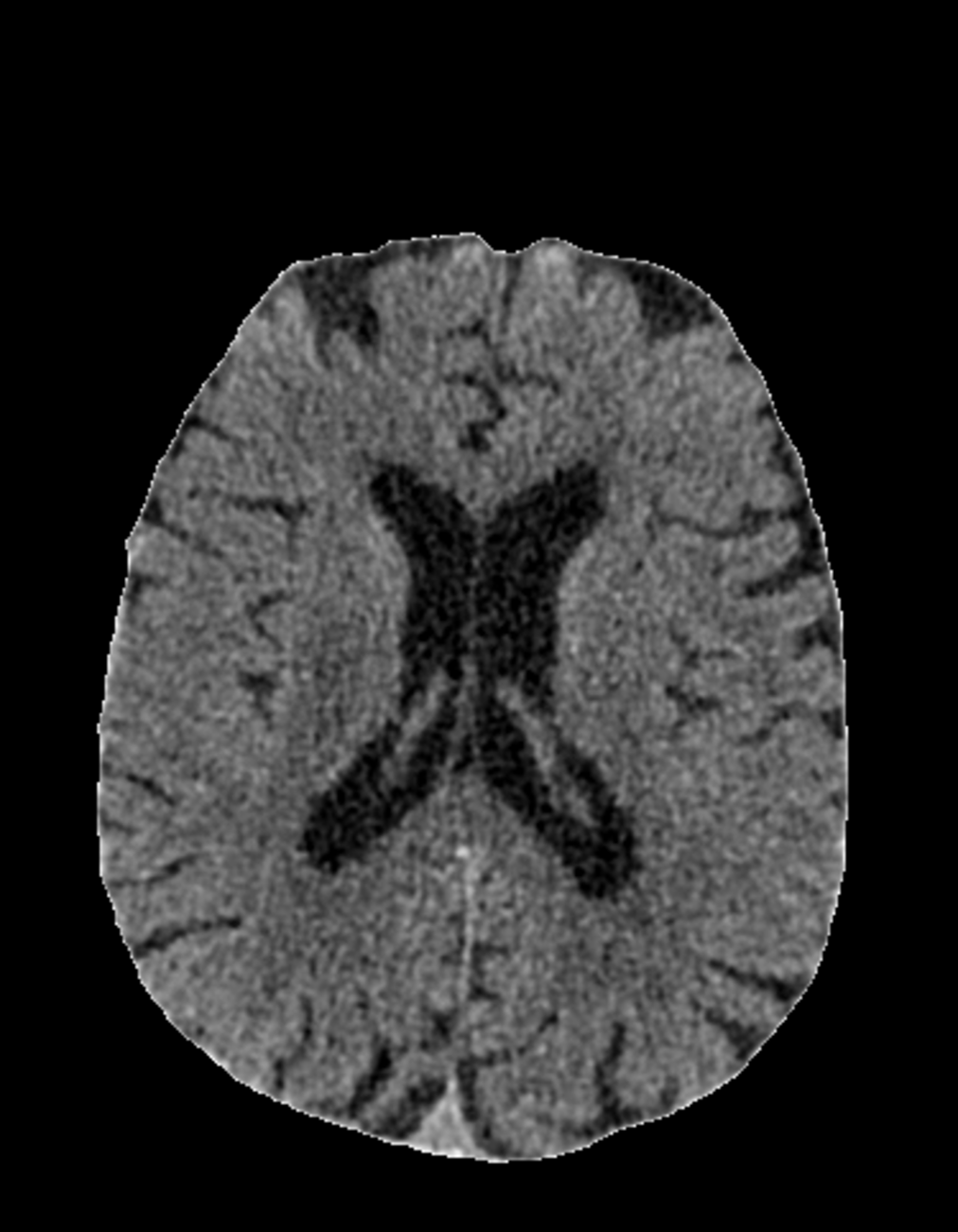} \\
        \textit{(a)} & \textit{(b)}
    \end{tabular}
    \caption{\textit{(a) CT image after the first pre-processing step, (b) after pre-processing}}
    \label{fig1}
\end{figure}

One crucial step in all deep-learning approaches is data pre-processing, which ensures data consistency. Training CNN directly on data without pre-processing leads to poor performance, as will be shown below. First, the intensities of CT images were thresholded between 0 and 80~Hounsfield units~\cite{46}. This transformation eliminates most of the artifacts and high-intensity tissues and remains visible such important parts of the brain as soft tissue, white matter, and gray matter. Second, the skull and coil were removed on each slice, leaving only the brain area on the CT image. This transformation was performed using connected component analysis. In particular, extraction of the largest connected component, then assigning zero values to pixels with the highest intensity, which correspond to the region of the skull, and re-extraction of the largest connected component. Third, each volume was normalized. We apply min-max normalization:
\begin{equation*}
    \widetilde{X} = \frac{X-X_{\min}}{X_{\max}-X_{\min}}
    \label{eq1}
\end{equation*}
where $X_{\min}$ and $X_{\max}$ are the minimum and the maximum intensity values ($X$) of 3D image, respectively. Such a normalization rescales values to~$[0,1]$. We also conducted an experiment applying standardization before min-max normalization:
\begin{equation*}
    \widehat{X} = \frac{X-\mu}{s}
    \label{eq2}
\end{equation*}
where $\mu$ is the mean and $s$ is the standard deviation of the brain region. Standardization was applied to each volumetric CT image of each patient independently. Also, the non-brain region was set to 0. Such a technique gives comparative intensity values in the brain area while invariant to the size of the background part. Fourth, the volumes were cropped to the non-zero region to dispose of a large uninformative background area. An example of a pre-processed image is shown in Figure~\ref{fig1}.

\begin{center}
\textbf{2.2. Patch-based approach}
\end{center}

The main difficulty related to 3D medical images is their large size, while computing resources are always limited. For example, a CT scan from our dataset has on average $512\times512\times405=106168320$~voxels (1.04~GB of memory for 32-bit voxels), where 405 is the average number of slices for a patient. It is worth noting that if the dataset is anisotropic, that is, the resolution of different axes of an image is not the same, resizing methods are not advisable. Different interpolation techniques can deform the physical object and remove small details. 

An optimal solution is to extract small parts, so-called patches, of 3D images and use them as input to a neural network. We use the uniform sampler, which extracts patches randomly from the whole volume with uniform distribution, and the weighted sampler, which extracts patches from different parts of the volume with a given non-uniform distribution. In our case, the weighted sampler extracts a patch with its center in the area of the healthy brain tissue with a probability of 0.5 and with its center in the area affected by the stroke with a probability of~0.5. The probability of the background as a patch's center is set to 0. During one training epoch, the same number of patches set to 32 are extracted for each patient. The patch size is set to $128\times128\times128$.

\begin{center}
\textbf{2.3. Neural network architecture}
\end{center}

\begin{figure}[h]
    \centering
    \includegraphics[width=\textwidth]{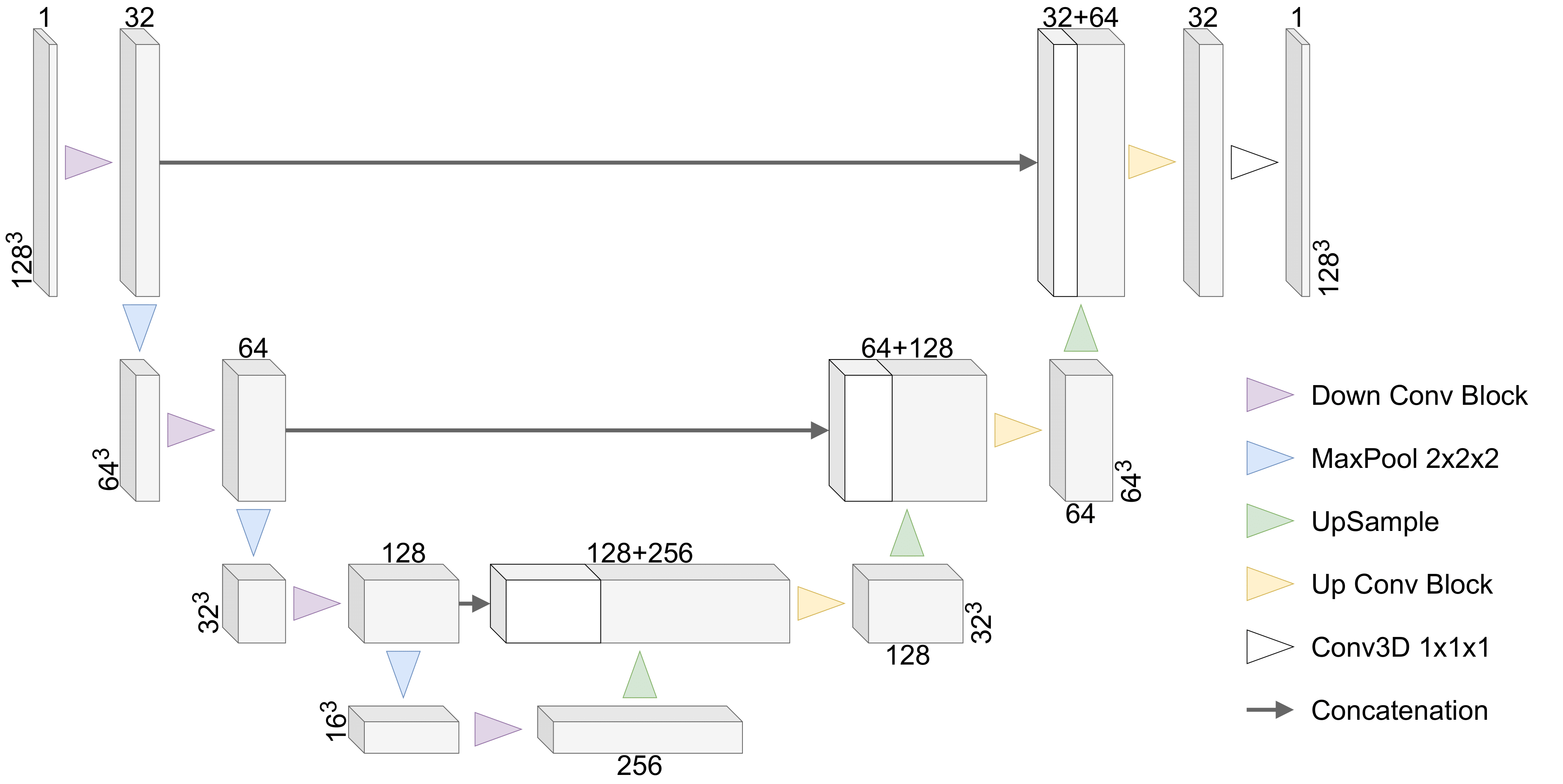}
    \caption{\textit{CNN architecture}}
    \label{fig2}
\end{figure}

Our deep learning algorithm builds on the encoder-decoder 3D U-Net~\cite{36} convolutional neural network, which we specially 
modify for our objective. The architecture of the neural network is shown in Figure~\ref{fig2}. The input size is $128\times128\times128$. Each convolutional block of encoder and decoder consists of two 3D convolutions with the size of kernel $3\times3\times3$ and stride of $1\times1\times1$, where each of them is followed by a normalization layer and activation function. We adopt LeakyReLU as an activation function since it showed better results in our experiments in comparison with more commonly used ReLU.

It is known that batch normalization~\cite{47} is strongly contingent on the current batch statistics during training. The statistics can include some noise, depending on the input examples. Therefore, it requires a sufficiently large batch size and also a large size of the training set. However, due to the high memory usage of 3D convolutions when employing large patches, we are limited to a maximum batch size of~2. Thus, we apply instance normalization, which solves this issue and shows better performance in our task than batch normalization.

Trilinear upsampling is used instead of the more traditional transposed convolution operator in the decoder part. In our research, we observe similar results, but the upsampling operator has no trainable parameters, so the occupied memory can be reduced. 

Residual connections mitigate the problem of vanishing and exploding gradients when training deep neural networks. It was first proposed in ResNet~\cite{48}. We integrate this technique in each convolution block of the contracting and expansive paths. The input is processed by 3D convolution with a kernel size of $1\times1\times1$, so the element-wise addition is possible. The architecture diagram of the convolutional block of the encoder with residual connection is illustrated in Figure 3. The parameters in brackets after Conv3D are the number of input channels, output channels, and kernel size. The decoder block is similar except for trilinear upsampling that halves the number of channels instead of 3D convolution that doubles the number of channels, therefore we leave aside the decoder block image.

We insert the Squeeze-and-Excitation (SE) mechanism~\cite{41} in our CNN as it shows strong performance in many computer vision tasks. It squeezes the global spatial information into a channel descriptor, captures inter-dependen\-cies of all channels, and then recalibrates the feature maps to accentuate relevant channels. In our case, it can help to learn where the affected area is located by strengthening the informative features and suppressing the weak ones.

\begin{figure}[h]
    \centering
    \includegraphics[width=.8\textwidth]{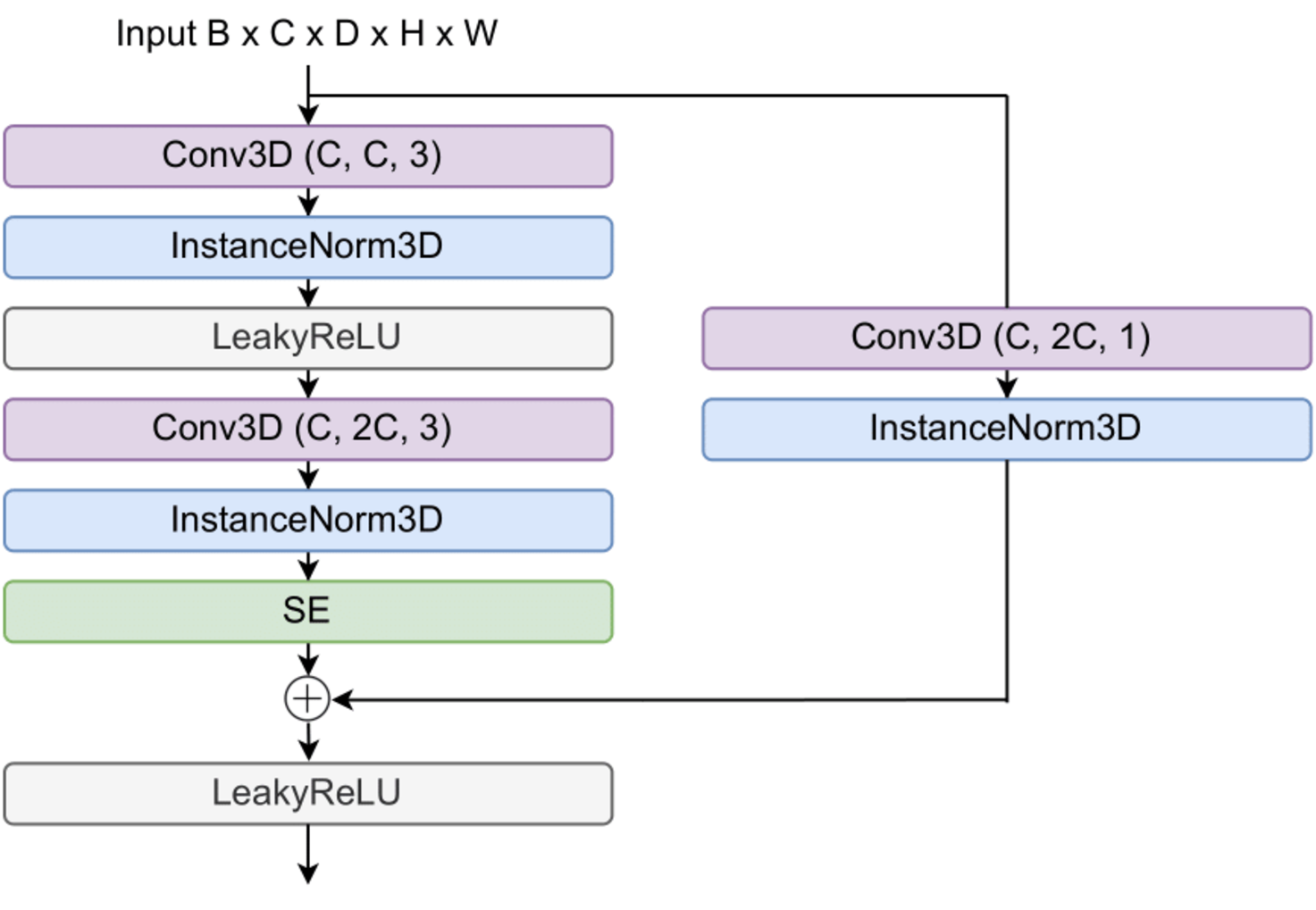}
    \caption{\textit{Convolutional block of the encoder}}
    \label{fig3}
\end{figure}

Channel-wise global average pooling is applied to the input tensor $T\in R^{C\times D \times H\times W}$ of the SE block. Then the obtained vector $U\in R^{C}$ is processed by the excitation mechanism: 
\begin{equation*}
    S = \sigma(W_2 \cdot \delta (W_1 \cdot U)), \quad S\in R^C
    \label{eq3}
\end{equation*}
where $W_1\in R^{\frac{C}{r}\times C}$ is a linear layer reducing the number of dimensions of the vector $U$, $\delta$ is the ReLU activation function, $W_2\in R^{C\times\frac{C}{r}}$ is a linear layer increasing the number of dimensions of the vector $U$, $\sigma$ is the sigmoid function, and $r$ is a parameter. In the end, a channel-wise multiplication between each element $s_i$ of the vector $S$ and the input tensor $T$ is performed. SE module is integrated into each convolutional block. Its particular location is shown in Figure~\ref{fig3}. 

\begin{center}
\textbf{2.4. Training procedure}
\end{center}

The obtained segmentation maps from the last CNN layer are transformed by the sigmoid function to get the probabilities of classes. Then the neural network weights are optimized using the sum of binary cross-entropy (BCE) and soft Dice~\cite{49} loss functions. We also perform experiments with weighted BCE loss since it addresses the class imbalance problem:
\begin{gather*}
    L = \sum_{i=1}^{N} w_i \left(-y_i \log p_i - (1 - y_i) \log (1-p_i)\right) + (1 - {\dfrac{2\sum\limits_{i=1}^{N} p_i y_i + \varepsilon}{\sum\limits_{i=1}^{N} p_i^{2} + \sum\limits_{i=1}^{N} y_i^{2} + \varepsilon}}),
    \\
    w_i = \dfrac{N_0y_i + N_1(1-y_i)}{N}
\end{gather*}
where $N_0$ is the number of background pixels of all training images, $N_1$ is the number of pixels related to the affected area, $N=N_0 + N_1$, $y_i$ and $p_i$ denote the ground truth and the confidence score of the model for pixel~$i$, respectively. 

\begin{center}
\textbf{2.5. Training, implementation and testing  details}
\end{center}

We train our networks using the following setting: the batch size of~2, Adam optimizer~\cite{50} with the initial learning rate of 1E-4, which is reduced by factor 2 every 5000 iterations until the learning rate is 1.25E-5, L2 weight decay of 1E-5. The reduction ratio $r$ of SE is~16. Each model was trained for a total of 40000 iterations.

The proposed algorithm was implemented in Python 3.7 using PyTorch 1.12.0 machine learning framework. We also employ TorchIO 0.18.80~\cite{51} Python library for data pre-processing. All trainings were conducted on NVIDIA Tesla T4 GPU with 16~GB of memory.

All patients were randomly split into five parts, so all experiments were performed using five-fold cross-validation. To evaluate our results, we measure Dice similarity coefficient (DSC), sensitivity, specificity, and precision metrics, which are most significant in medical image segmentation. The use of these metrics, in addition to the main metric DSC, allows us to evaluate aspects of the behavior of segmentation methods in conditions of unbalanced samples. Results are presented as mean $\pm$ standard deviation. DSC is calculated similarly to the soft Dice loss but is not subtracted from 1 and thresholded values are used instead of confidence scores. To get the predicted binary segmentation mask, the obtained probabilities were binarized according to the threshold of~0.5. On the test set, patches were extracted across a whole volume with an overlap of 25\% over a grid. The predictions in the overlapping areas were averaged.

\begin{center}
\textbf{3. Experiments and results}
\end{center}

\begin{table}[!b]
    \caption{\textit{Cross-validation results}}\label{tab1}
    \centering\small
    \tabcolsep=0.7mm
    \begin{tabular}{|l|c|c|c|c|}
        \hline
        \thead{Method} & DSC & Sensitivity & Specificity & Precision \\\hline
        No pre-processing & $0.508\pm0.037$ & $0.597\pm0.09$ & $0.9975\pm0.0014$ & $0.519\pm0.057$ \\\hline
        Pre-proccesing & $0.511\pm0.023$ & $0.518\pm0.035$ & $0.9980\pm0.00092$ & $0.617\pm0.038$ \\\hline
        \makecell[l]{Pre-processing +\\weighted sampler} & $0.547\pm0.025$ & $0.567\pm0.076$ & $0.9975\pm0.0016$ & $0.616\pm0.069$ \\\hline
        \makecell[l]{Pre-processing +\\weighted sampler +\\residual connections} & $0.574\pm0.016$ & $0.606\pm0.064$ & $0.9975\pm0.0013$ & $0.615\pm0.049$ \\\hline
        \makecell[l]{Pre-processing +\\weighted sampler +\\residual connections\\+ SE} & $0.612\pm0.064$ & $0.689\pm0.059$ & $0.9965\pm0.0020$ & $0.619\pm0.044$ \\\hline
        \makecell[l]{Pre-processing +\\standardization +\\weighted sampler +\\residual connections\\+ SE +\\weighted loss} & $0.628\pm0.033$ & $0.699\pm0.039$ & $0.9965\pm0.0016$ & $0.619\pm0.036$ \\\hline
    \end{tabular}
\end{table}

The quantitative comparison of our methods is presented in Table~\ref{tab1}. The pre-processing of the first experiment included only thresholding of the Hounsfield units and the min-max normalization. Also, patches were derived using the uniform sampler. The results of the second experiment show that adding the robust pre-processing techniques, including selecting the brain tissue and cutting to the non-zero region, increases DSC by 0.4\%. The restricted patch extraction method performed by the weighted sampler during training improves performance by 3.6\% of DSC and 4.9\% of sensitivity. The main proposed modifications of 3D U-Net are residual connections and SE. While the integration of the residual connections improves the DSC by 2.7\%, the consistent inserting of the SE modules gives a 3.8\% of DSC increase, also showing the promising sensitivity value of $0.689\pm0.059$. The final algorithm configuration maintains the above-described improvements, while also standardization before min-max normalization and training using the weighted loss function are performed. Such a configuration gives the best results, $0.628\pm0.033$ of DSC and $0.699\pm0.039$ of sensitivity.

We also study how the patch overlapping size impacts the segmentation accuracy when performing testing. Table~\ref{tab2} shows the quantitative assessment with 25\%, 50\%, and 75\% path overlapping using the final algorithm configuration. Although the overlapping of 75\% gives the best results relying on DSC, it reduces sensitivity, and the running time of the algorithm per one patient is increased by 10~times.

We tried to insert in the training process various augmentation techniques, such as random affine transformation, random flip, random elastic deformation, random gamma intensity transformation, but it worsened the performance of our algorithm.

\begin{table}[!t]
\caption{\textit{The comparison of the patch overlap sizes}}
    \centering\small
    \tabcolsep=1.2mm
    \begin{tabular}{|c|c|c|c|c|c|}
        \hline
        \makecell[c]{Overlap\\size, \%} & \makecell[c]{Time per\\patient, s} & DSC & Sensitivity & Specificity & Precision \\\hline
        25 & 45 & $0.628\pm0.033$ & $0.699\pm0.039$ & $0.9965\pm0.0016$ & $0.619\pm0.036$ \\\hline
        50 & 90 & $0.622\pm0.022$ & $0.693\pm0.047$ & $0.9965\pm0.0016$ & $0.629\pm0.036$ \\\hline
        75 & 440 & $0.630\pm0.025$ & $0.693\pm0.049$ & $0.9965\pm0.0016$ & $0.638\pm0.030$ \\\hline
    \end{tabular}
    \label{tab2}
\end{table}

\begin{figure}[!b]
    \centering
    \includegraphics[width=.95\textwidth]{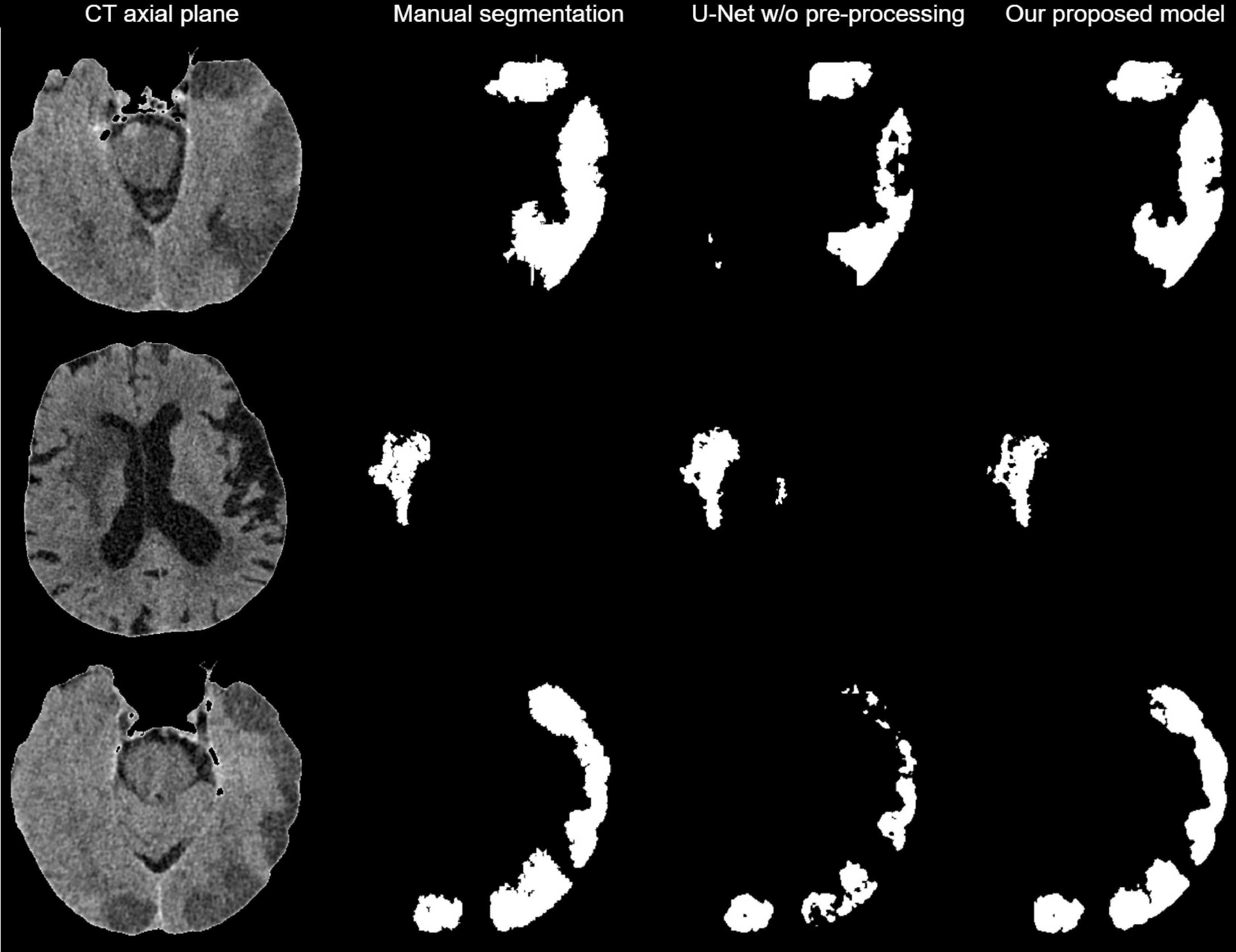}
    \caption{\textit{Qualitative segmentation results}}
    \label{fig4}
\end{figure}

Qualitative results are presented in Figure~\ref{fig4}. Several axial plane slices of the 3D images are from the validation set.

\begin{center}
\textbf{Conclusion}
\end{center}

In this work, we presented and evaluated an automatic algorithm for the segmentation of acute ischemic stroke lesion in non-contrast computed tomography brain images. Our deep learning approach is based on the 3D~U-Net convolutional neural network. As far as we know, no previous research has investigated the combination of 3D~U-Net, Squeeze-and-Excitation blocks, residual connections, specific patch extraction technique, and weighted loss function for solving the problem of acute ischemic stroke lesion segmentation on NCCT images. We have experimentally shown that the suggested pipeline for NCCT images segmentation, as compared to other state-of-the-art methods, improved segmentation performance by implementing robust pre processing techniques, Squeeze-and- Excitation blocks, and residual connections. Also, specific patch extraction technique and weighted loss function partially solved the class imbalance problem and increased the segmentation accuracy. Our proposed model showed an average Dice of $0.628\pm0.033$, sensitivity of $0.699\pm0.039$, specificity of $0.9965\pm0.0016$, and precision of $0.619\pm0.036$. High specificity values are caused by the dominance of the negative class (non-affected tissue) over the positive class (pathological area) in the sample. The method can be used by radiologists in delineating between damaged and healthy brain tissue and deciding on further treatment. In particular, it can help correct inaccuracies in their stroke area predictions. Moreover, the method can assist doctors with the large flow of patients by selecting cases with affected brain areas from normal ones. Thus, doctors first of all examine patients in need of emergency care.

It is also important to note that one of the main constraints of our objective is the small amount of data in our dataset. It is with this we associate the unusual obtained values of the specificity metric, and with the class imbalance problem. In the future, we plan to explore other CNN architectures and increase the dataset.

\begin{center}
\textbf{Acknowledgments}
\end{center}

The work was partly supported by RFBR grant No.~19-29-01175, and by the State Contract of the Sobolev Institute of Mathematics, Project No. FWNF-2022-0015.

\begin{center}
\textbf{Authors’ information}
\end{center}

\textbf{Anna Vadimovna Dobshik} (b. 1999), bachelor of applied mathematics and informatics, graduated from Novosibirsk State University, Department of Mechanics and Mathematics in 2021. Works as a machine learning engineer for DeepSound AI. Research interests are computer vision, medical image segmentation.  E-mail: \underline{\textit{a.dobshik@alumni.nsu.ru}}.

\textbf{Sergey Konstantinovich Verbitskiy} (b. 1999), bachelor of applied mathematics and informatics, graduated from Novosibirsk State University in 2021. Works as a machine learning engineer at StyleDNA. Research interests are image segmentation, signal processing, speech recognition, object detection, and video action recognition. Email:
\underline{\textit{s.verbitskii@alumni.nsu.ru}}.

\textbf{Igor Alekseevich Pestunov} (b. 1955), graduated from Novosibirsk State University in 1977. PhD in physics and mathematics (1998). He is a key researcher at Federal Research Center for Information and Computational Technologies. His current research interests include clustering, pattern recognition, and image analysis. Email: \underline{\textit{pestunov@ict.nsc.ru}}.

\textbf{Kseniya Mihailovna Sherman} (b.1989), graduated from Novosibirsk State University in 2012. Residency in traumatology and orthopedics; specialization in radiology until 2017. Currently works as a radiologist at the International Tomography Center. Scientific interests: introduction of new technologies into medical practice. E-mail: 
\underline{\textit{Ksh1420@yandex.ru}}.

\textbf{Yuriy Nikolaevich Sinyavskiy} (b. 1983), graduated from the Novosibirsk State University in 2005. PhD in Technical Sciences (2021). Currently he works as the researcher at the Federal Research Center for Information and Computational Technologies. Research interests are image segmentation, geoinformatics, and software development. E-mail: \underline{\textit{yorikmail@gmail.com}}.

\textbf{Andrey Alexandrovich Tulupov} (b. 1981), graduated from the Moscow State University in 2005. PhD in Medical Sciences (2006). Currently he works as the chief researcher at The Institute International Tomography Center of the Russian Academy of Sciences. Research interests are magnetic resonance imaging and computed tomography. E-mail: 
\underline{\textit{taa@tomo.nsc.ru}}.

\textbf{ Vladimir Borisovich Berikov} (b. 1964), graduated from Novosibirsk State University in 1986. Received candidate’s degree in 1996 and doctoral degree in 2007. Works at Sobolev Institute of Mathematics SB RAS, acts as a head of laboratory and chief researcher. Research interests: mathematical methods of data analysis and their application in image processing and medical data. Email: \underline{\textit{berikov@math.nsc.ru}}.

\end{document}